# Automatic segmentation and determining radiodensity of the liver in a large-scale CT database

Nikolay S. Kulberg[1,3], Alexey B. Elizarov[1], Vladimir P. Novik[1], Victor A. Gombolevsky[1], Anna P. Gonchar[1], Amel L. Alliua[2], Victor Yu. Bosin[1], Anton V. Vladzymyrsky[1], Sergey P. Morozov[1]

[1] State Budget-Funded Health Care Institution of the City of Moscow "Research and Practical Clinical Center for Diagnostics and Telemedicine Technologies of the Moscow Health Care Department", Moscow, Russia.

[2] Federal State Budgetary Scientific Institution "Russian Scientific Center of Surgery named after Academician B.V. Petrovsky", Moscow, Russia.

[3] Federal Research Center "Computer Science and Control" of Russian Academy of Sciences, Moscow, Russia

E-mail: v.novik@npcmr.ru. (Corresponding author)

**Abstract.** This study proposes an automatic technique for liver segmentation in computed tomography (CT) images. Localization of the liver volume is based on the correlation with an optimized set of liver templates developed by the authors that allows clear geometric interpretation. Radiodensity values are calculated based on the boundaries of the segmented liver, which allows identifying liver abnormalities. The performance of the technique was evaluated on 700 CT images from dataset of the Unified Radiological Information System (URIS) of Moscow. [http://medradiology.moscow/eris]. Despite the decrease in accuracy, the technique is applicable to CT volumes with a partially visible region of the liver. The technique can be used to process CT images obtained in various patient positions in a wide range of exposition parameters. It is capable in dealing with low dose CT scans in real large-scale medical database with over 1 million of studies.

# Introduction

Abnormal of liver development and diseases of the hepatobiliary system rank third in the prevalence rate in the population after bronchopulmonary lesions and cardiovascular diseases [1, 2, 3]. To diagnose liver diseases, it is required to determine its geometric characteristics and densitometric parameters. For example, the diagnosis of fatty liver disease is based on determining the radiodensity of the liver [4, 5]. This is a routine process that can be time-consuming. Automatic methods for segmentation and determining radiodensity of the liver will significantly optimize clinician workflow by performing appropriate calculations quickly and with high accuracy. Automatic data analysis will also allow identifying subclinical liver diseases during CT studies of other organs (including screening) when the scan area partially covers the liver.

Liver segmentation can be performed by semi-automatic and automatic methods. Semi-automatic methods [6] are time- and effort-consuming due to interactive operations. But there is number of studies aimed at reducing the expert efforts to a limited number of actions such as seed point selection.

Automatic detection of the liver volume in CT images from a large database can be difficult task as CT scans are obtained in different clinical settings on a variety of scanners. In cases where the lungs are examined, the liver volume is partially present and noisy. The contrast of the boundaries between the liver and neighboring structures is unstable. Given that, developing a reliable algorithm to deal with CT images is challenging task.

For more than 40 years of image processing history in CT, numerous methods for liver segmentation have been developed. There are several thorough reviews in this field [7, 8]. The probabilistic atlas method uses a map that describes the probability of voxel with specific coordinates to constrain the object of interest. To form a probabilistic map of the liver, a set of CT images with expert labeled liver segmentation is used [9]. In [10], the authors use the probabilistic atlas approach and classification of voxels into the "liver" and "not-liver" classes. Papers [11, 12] use shape-intensity prior level set combining probabilistic atlas and probability map constrains. Statistical Shape Model [13, 14, 15, 16] allows estimating the liver volume based on the ground truth marking in a trained database. To improve accuracy, deformable models were used to locally adapt the liver shape to a precise boundary [17, 18]. Graph Cut method describes cropped image volume as a graph and finds an optimal partition of the graph to solve the max-flow problem. [19, 21, 22] successfully use this approach. Region-growing approach for automatic detection is applied in [23]. Neural network methods in liver segmentation have been employed in [24] since 1994. Convolution Neural Networks (CNN) have been used recently to achieve state-of-art results in automatic liver segmentation. Currently, they produce highly competitive results comparable to state-of-the-art methods [15].

Despite a large number of studies in this area, the available algorithms have insufficient accuracy or reliability in case of partially visible liver, which is very important when dealing with images in large databases.

Thus, at the moment, there is no ready-made solution that allows segmenting the liver in automatic mode in a large number of cases of practical interest. The objective of this work is to develop a technology for automatic segmentation of the liver and automatic determination of its radiodensity based on CT data, which can be applicable to cases of pathological changes of the liver when covering the liver volume partially or completely. Partial cover cases are usual for chest CT scans.

## Methods

The authors have developed a technology for automatic segmentation and determining radiodensity of the liver. The automatic segmentation algorithm consists of the following steps:

1. Determination of the anatomical region of the tomogram by analyzing skeleton bones in the CT image. This step allows to exclude cases where the tomogram does not contain the liver, even partially.

2. Determination of the approximate position and boundaries of the liver using a correlation algorithm with the optimal template from the base of liver shape templates, and the position and size adjustment of the template in the CT volume.

3. Determination of densitometric characteristics of the liver by the boundaries found.

4. Determining of accurate liver boundaries by an algorithm that uses the obtained densitometric characteristics and approximate liver boundaries.

To determine the anatomical region, a correlation algorithm is used. It matches bone tissues in the CT image and a skeleton template. The skeleton template is obtained by processing a tomogram containing a full-length image of an adult. This step of the algorithm allows to exclude images where the liver is absent, and to avoid significant errors in the next steps of the algorithm. Fig. 1 shows examples of determining different anatomical areas. In the processed CT image, the algorithm determines the approximate position of an object similar to the skeleton template. Thus, when processing a tomogram that does not contain the liver (e.g., head CT), based on the found position, the system decides not to continue searching for the liver on the tomogram. But if the tomogram position is consistent with the location of the liver on the template, the algorithm proceeds to the next step, i.e. determination of the liver location based on the correlation of the organ shape with the shape templates.

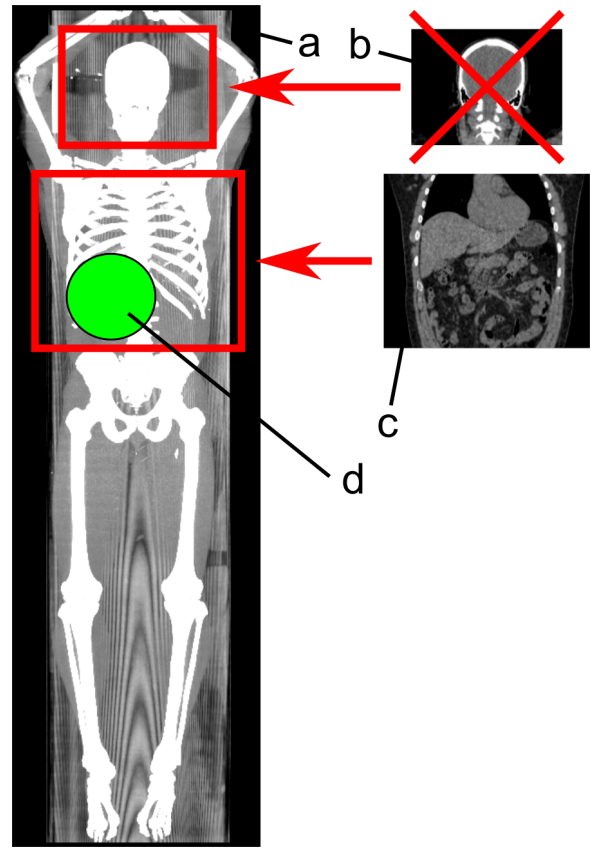

**Fig. 1. Matching a CT image with the skeleton template. Excluding the cases when the CT area does not capture the liver. a—CT image containing the full skeleton of an adult; b—CT image representing the head; c—CT image representing abdominal cavity; d—area of the liver on the template**

It is reported [26], that the liver is quite variable in shape and size. Irregular geometric shape of the liver as well as the presence of a wide range of its individual variants determines the complexity in calculating its volume. There is a considerable variation in the definition of liver shapes. Classification methods developed in the "pre-computer" era are based on analyzing the ratio of longitudinal and transverse dimensions of the liver, the ratio of sizes of its lobes or on selection of characteristic forms of the liver. However, a unified classification system has not been developed.

The widespread use of computer technology allowed to create more complex liver shape models. However, this has not result in a clear classification of the liver shapes. Rather, on the contrary, the idea of unpredictability of the liver shape has emerged. E.g., in work [27], the classification of liver forms is based on the expansion of the segmented liver surface in a spherical function series with subsequent analysis of the series coefficients. The result of classification is a combination of 16 coefficients, which actually means almost infinite variability of the forms. A visual interpretation of this classification is difficult.

Thus, there is no generally accepted classification of the liver shape in modern literature. In this work, the authors have developed a simple classification scheme that allows clear geometric interpretation (Fig. 2). This scheme can be considered as a reduced model presented in [9]. Six types of the liver were identified based on linear dimensions of the right and left lobes of the liver:

Type I: normal sizes of the right and left lobes;

Type II: normal right and elongated left lobe;

Type III: normal right and shortened left lobe;

Type IV: elongated right and normal left lobe;

Type V: elongated right and left lobes;

Type VI: elongated right and shortened left lobe.

The judgment about the type of the right lobe (shortened, normal, elongated) was based on the craniocaudal size, the normal dimensions of which are within 13.5-15.5 cm [29]. The left lobe of the liver was evaluated visually.

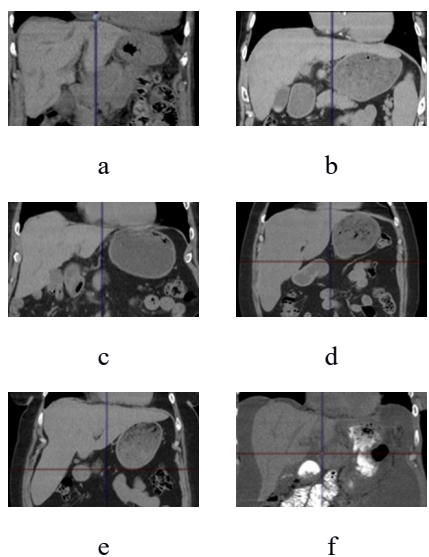

**Fig. 2. Types of the liver shape: a—type I: normal sizes of the right and left lobes; b—type II: normal right lobe and elongated left lobe; c—type III: normal right lobe and shortened left lobe; d—type IV: elongated right lobe and normal left lobe; e—type V: elongated right and left lobes; f—type VI: elongated right lobe and shortened left lobe**

For each of these types, two characteristic tomograms were selected. On their basis, in manual mode, three-dimensional binary liver masks were created by cross-sections (Fig. 3) and a base of three-dimensional liver shape templates was formed (12 templates in total, used in step 2 of the automatic segmentation algorithm).

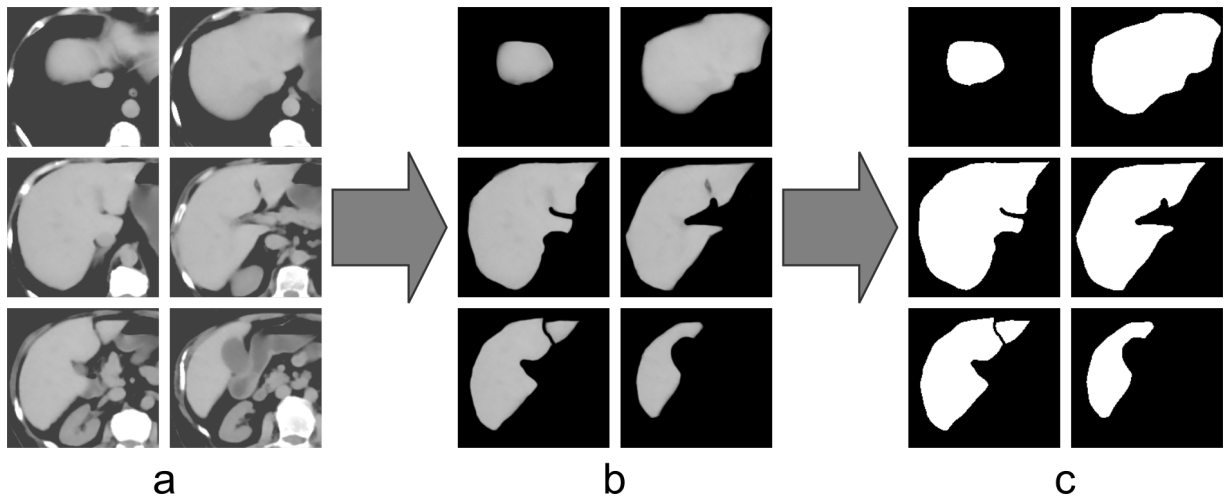

**Fig. 3. Formation of a three-dimensional liver template by cross-sections:**
**a—axial CT image of the liver; b—manually marked ROI of the liver;**
**c—binary mask of the liver**

When determining the approximate position and boundaries of the organ, a correlation search of the liver is performed using all the 12 templates. Some ideas of the proposed technique are close to those mentioned in [13]. Three-dimensional liver template is used, rather than with its particular sections, as it is done in many other works. The template scale also varies during the search. As a result, the system determines the template that best fits the 3D image of the liver on the tomogram, as well as the optimal position and scale of this template.

In general, procedures performed in steps 1 and 2 repeat the actions performed by the doctors when analyzing a tomogram: first, the anatomical region of the tomogram is assessed as a whole; and then a visual search for the organ of interest is performed based on the analysis of object shapes and anatomical landmarks.

To accurately determine liver boundaries in a CT scan, it is planned to develop an algorithm that uses clustering (similar to the principal components method) and geometric connectivity criteria [30]. It will use the densitometric characteristics calculated by the found approximate liver boundaries (Fig. 4).

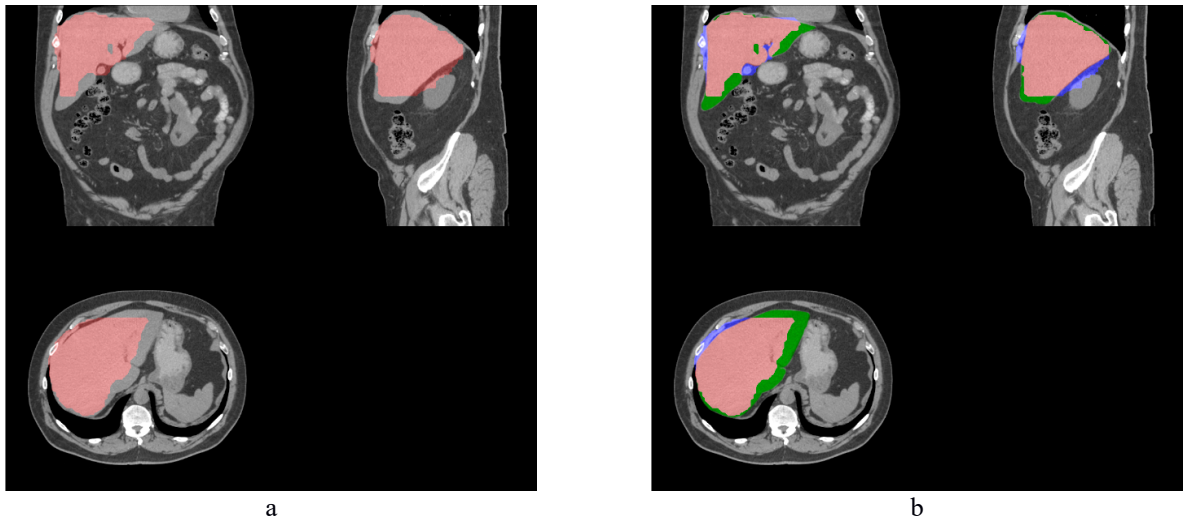

a                                                                                b

**Fig. 4.** Accurate adjustment of liver boundaries position: a—initial position of the boundaries (according to the template matching); b—adjusted position of the boundaries; red color—template area; blue color—excluded areas; green color—added areas

The developed algorithm is applicable for a great variety of densitometric characteristics of the liver as it operates without *a priori* hypotheses on the density of the liver and surrounding organs. Due to this, it correctly localizes the liver with significant density deviations from the norm. The algorithm takes into account a possible heterogeneity of the organ. If the analyzed volume contains multiple areas with different density values (normal tissue and pathological lesions), density metrics are calculated separately for each area.

The technology was tested on CT data obtained from the Unified Radiological Information System of Moscow. For testing, tomograms of patients with normal liver and those with various liver pathologies were used. The authors used abdominal and chest CT scans where the liver is not always completely visible. Considering that chest CT is the most common CT exam, the situations where the liver is partially present in a CT scan are typical. The proposed technique allows checking if the CT scans can be used to detect subclinical liver diseases.

To evaluate the specificity of liver segmentation, the developed technology was also tested on data that do not contain liver images (tomograms of the head, neck, and extremities). In total, 481 CT images were used to determine the sensitivity of the technology, and 316 CT images were used to determine its specificity.

## Results

The segmentation technique was tested on 797 CT images including both low-dose and normal-dose studies. Anonymized images were obtained from large-scale medical database URIS. The sensitivity of the technology is 95.6%, the specificity is 100% and the AUC is 0.978. The segmentation algorithm correctly detects the position of the liver under challenging conditions, where:

1) The CT image includes at least 40% of the liver volume (in case of the chest CT);
2) Liver density differs significantly from the normal values (various pathological changes in the liver).

To analyze the accuracy of determining the liver density, the authors have randomly selected 89 CT scans of patients aged 50 to 74 years. When measured manually, the liver density values range from -4.9 HU to 72.6 HU. The standard deviation of the density values when measured automatically as compared to manual measurement is 4.3 HU, the 95th percentile is 5.8 HU. The maximum density deviation is 17 HU. Such a large maximum deviation is caused by segmentation errors on some tomograms. 99.5% of studies lie within the 10 HU range, such accuracy is acceptable for screening applications.

As shown by the test results, the technology is applicable to tomograms of both normal and pathological liver with significant deviations of radiodensity from the normal values due to the presence of pathological tissues of large size and can be used for automatic analysis of CT scans obtained in examinations of other organs where the CT volume partially covers the liver. The developed segmentation algorithm shows high robustness, the radiodensity of the liver is determined with high accuracy. The accuracy is satisfactory for liver pathologies screening by processing large databases.

The authors have implemented the developed technique in C++. The data processing framework allows downloading CT images from a remote storage and processing them on a local workstation. The average processing time of a typical CT image with a size of 512×512×360 is about 50 seconds for desktop system with 3 GHz 4-core Intel Core i5 processor and 8 GB DDR3 memory.

The only solution available to the vast majority of clinicians is the CT Liver Analysis application developed by Philips as part of the IntelliSpace Portal software [https://www.philips.co.uk/healthcare/education-resources/publications/hotspot/isp-ct-liver-analysis]. The system automatically detects the liver in CT scans. In a semi-automatic mode, clinicians can segment the liver into lobes by marking anatomical landmarks. However, this system sometimes fails to detect the liver if its radiodensity differs significantly from the normal range (the normal range is from +40 HU to +60 HU). Moreover, this system cannot process tomograms containing an incomplete image of the liver (such as chest CT scans); however, such cases are of great interest in the automatic analysis of large datasets as they allow detecting subclinical liver diseases. Note that chest is the most common CT exams. During our research, we have found several cases when the CT Liver Analysis software fails. It happens more frequently in cases where the liver is partially visible. As compared to the IntelliSpace Portal software tools (Philips), the proposed technology gives more stable results of automatic liver detection. For example, in case of a liver neoplasm of a large volume (density of +23.5 HU, which differs significantly from the normal liver density), the IntelliSpace Portal software excluded the neoplasm from the liver. At the same time, the system developed by the authors segments the liver together with the neoplasm, which enables automatic density determination of the normal tissue and pathological lesions of the liver in the sample.

Fig. 5 and 6*a* show automatic liver segmentation result in this sample as obtained with the IntelliSpace Portal software (Philips) and the proposed system. Fig. 6b and *c* show the results of automatic determining densitometric characteristics of the liver in the given sample produced by the developed system.

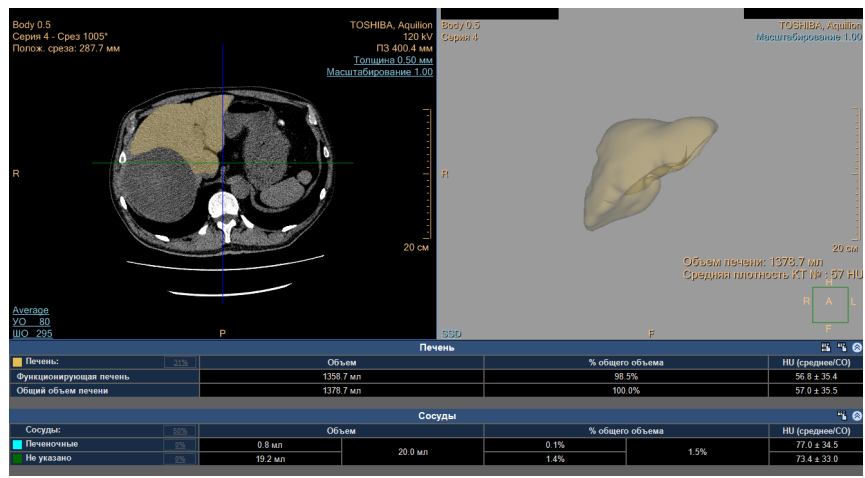

**Fig. 5. Automatic liver segmentation with IntelliSpace Portal software (Philips) on the sample with a liver neoplasm of a large volume**

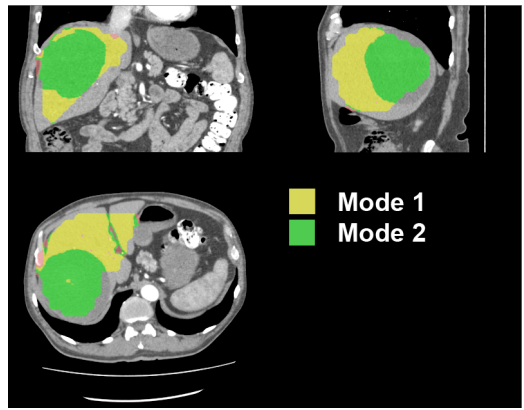

*a*

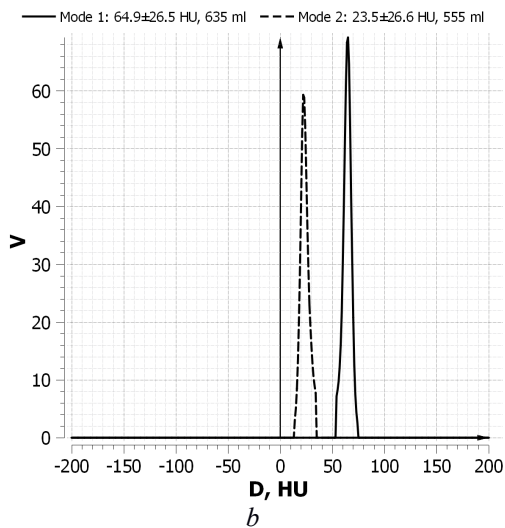

*b*

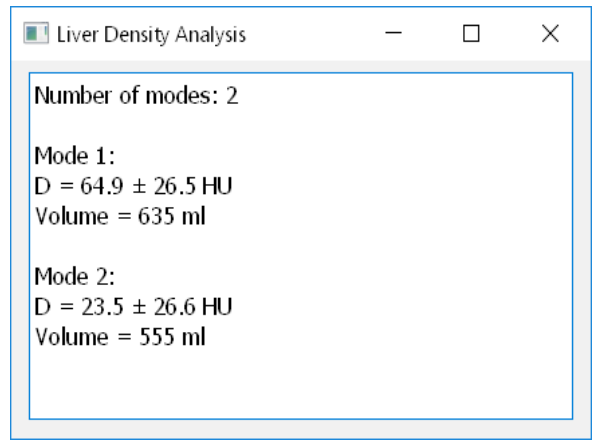

*c*

**Fig. 6. Automatic liver segmentation and determining the density of normal and pathological tissues by the developed system on the sample with a liver neoplasm of a large volume: a—segmentation result, areas of different density marked with different colors; b—distribution of radiodensity values with selected areas of different density (modes); c—automatically generated text report on determining the volume and radiodensity**

Another example represents the liver with a significantly reduced density of +13.7 HU. In this case, the IntelliSpace Portal software failed to detect the liver (Fig. 7), while the developed system successfully segmented it (Fig. 8). Processing such data is of great practical and scientific value for the healthcare.

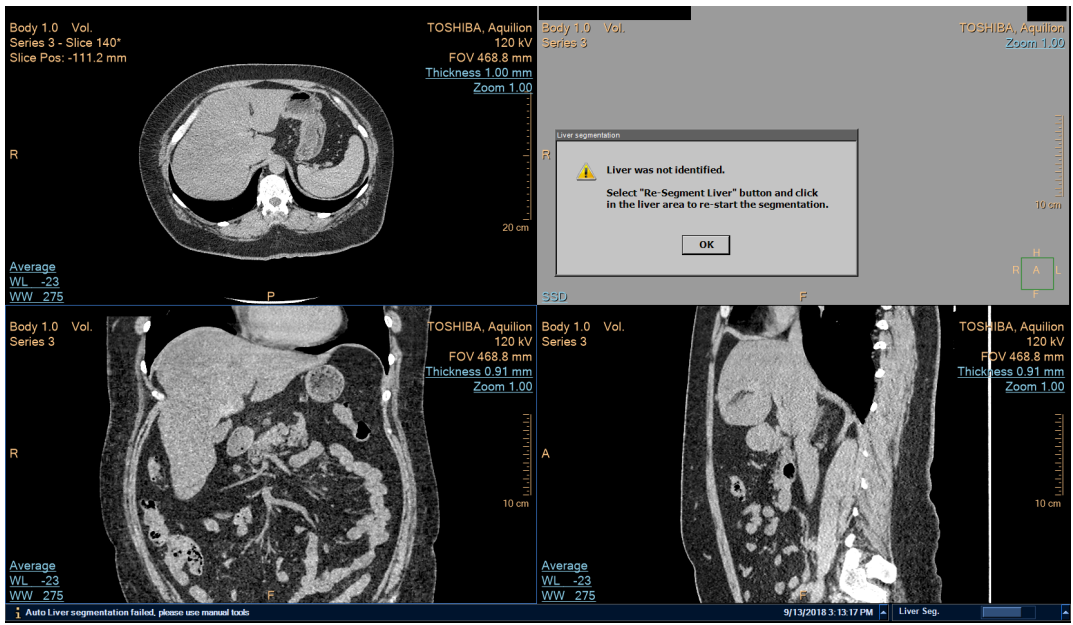

Fig. 7. Automatic liver segmentation with IntelliSpace Portal software (Philips) on the sample with a significantly reduced liver density: no liver detected

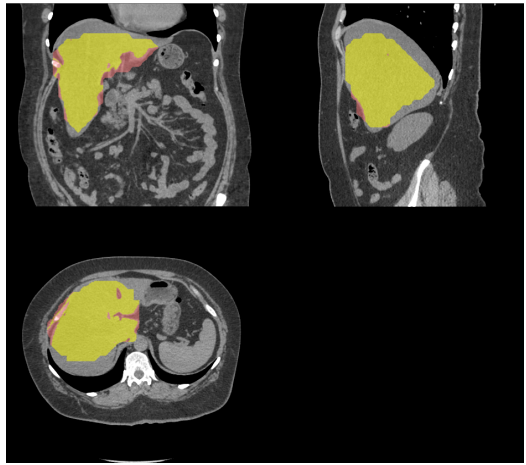

*a*

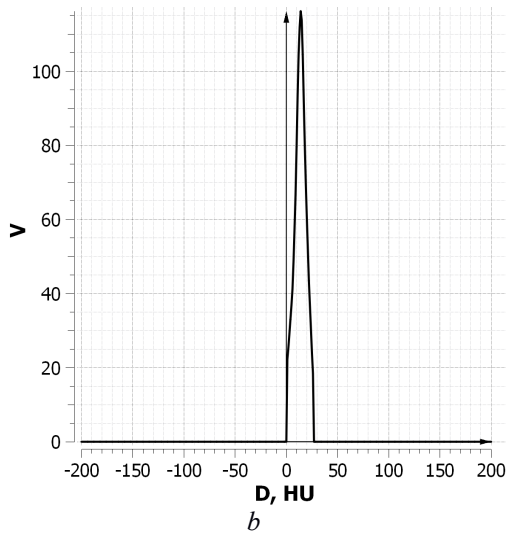

*b*

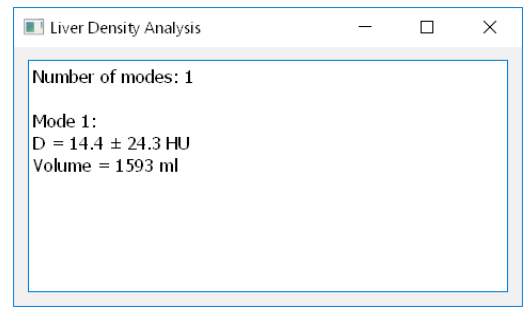

*c*

**Fig. 8. Automatic liver segmentation and determining the liver tissue density by the developed system on the sample with a significantly reduced liver density: a—segmentation result; b—distribution of radiodensity values; c—automatically generated text report on determining the volume and radiodensity**

To improve technology performance, the authors intend to make a more careful selection of the liver shape templates. To reduce the influence of the noise on the calculated densitometric parameters, algorithm for filtering CT data is applied to CT images [31].

## Conclusions

The developed technology relieves the radiologist from routine liver segmentation and densitometry tasks. It enables automatic analysis of large databases of CT images to detect subclinical liver diseases and to perform medical research based on liver masks.

Overall, 481 CT images were used to determine the sensitivity of the technology, and 316 CT images were used to determine its specificity. The accuracy is satisfactory for screening liver pathologies by processing large databases. The sensitivity of the technology is 95.6%, the specificity is 100% and the AUC is 0.978. The segmentation algorithm correctly detects the position of the liver under challenging conditions:

Authors contribution in the article includes 3 points:
- Automatic determination of the liver boundaries;
- Automatic determination of densitometric characteristics of the liver;
- Automatic analysis of CT scans obtained in clinical settings and stored in databases to diagnose subclinical liver diseases;

The proposed technique is able to process CT images where the liver volume is partially present. This property is quite unique for a liver segmentation algorithm. The authors have tested the developed technique on CT images from a real large-scale database maintained on the basis of pulmonary medical centers of Moscow. The developed technique allows detecting CT images of patients with subclinical liver diseases. It is also planned to use the technology for automatic segmentation and determining the densitometry characteristics of other abdominal organs.

## Declarations

*Ethics approval and consent to participate*

*Consent for publication*

*Availability of data and material*

*Competing interests*

The authors declare that they have no competing interests.

*Funding*

The research is supported by the Russian Foundation for Basic Research, grant #17-01-00601.

*Authors' contributions*

NSK was in charge of the technical part of the work and participated in writing the article.

ABE developed data processing algorithms and participated in writing the article.

VAG formulated the clinical problem and participated in writing the article.

NVP participated in discussion about methods to use and in writing the article.

APG performed data processing and clinical evaluation of processing results.

ALA analyzed the types of liver forms.

VYB provided the clinical rationale for the feasibility of work and participated in writing the article.

AVV was in charge of the clinical part of the work and conducted a statistical evaluation of processing results.

SPM provided overall project management.

All authors read and approved the final manuscript.

*Acknowledgements*

The research is supported by the Russian Foundation for Basic Research, grant #17-01-00601.